\documentclass[draft,preprint,aps,eqsecnum,amsmath,amssymb]{revtex4}
\begin{document}

\newcommand{\beq}{\begin{equation}}
\newcommand{\eeq}{\end{equation}}
\newcommand{\bea}{\begin{eqnarray}}
\newcommand{\eea}{\end{eqnarray}}
\newcommand{\cir}{{\buildrel \circ \over =}}

\title{Canonical Reduction of Gravity: from General Covariance to
Dirac Observables and post-Minkowskian Background-Independent
Gravitational Waves.}

\author{Luca Lusanna}

\affiliation{Sezione INFN di Firenze\\Polo Scientifico, via Sansone 1\\
 50019 Sesto Fiorentino, Italy\\
 E-mail LUSANNA@FI.INFN.IT}

\begin{abstract}
The status of canonical reduction for metric and tetrad gravity in
space-times of the Christodoulou-Klainermann type, where the ADM
energy rules the time evolution, is reviewed. Since in these
space-times there is an asymptotic Minkowski metric at spatial
infinity, it is possible to define a Hamiltonian linearization in
a completely fixed (non harmonic) 3-orthogonal gauge without
introducing a background metric. Post-Minkowskian
background-independent gravitational waves are obtained as
solutions of the linearized Hamilton equations.
\bigskip

Talk given at the Symposium QTS3 on Quantum Theory and Symmetries,
Cincinnati,~September~10-14 2003.
\end{abstract}

\maketitle


In general relativity the metric tensor has a double role: it is
the potential of the gravitational field and simultaneously
describes the chrono-geometrical structure of the space-time in a
dynamical way by means of the line element. It teaches to all the
other fields relativistic causality: for instance it selects the
paths to be followed by the rays of light in the geometrical optic
approximation. This aspect of the metric tensor is completely lost
every time it is split in a background metric plus a perturbation.
This implies that, like in special relativity, we introduce a
background non-dynamical (absolute) chrono-geometrical structure
and the perturbation (like every other massless field) propagates
along the fixed background light-cone. As a consequence, since we
do not know realistic solutions of Einstein's equations for
macroscopic bodies, the only existing theory of measurement in
general relativity is the axiomatic one of Ehlers, Pirani and
Schild \cite{1}, which utilizes {\it test} massive particles and
rays of light in place of their {\it dynamical} counterparts.
Moreover, we have that in the approaches to quantum gravity based
on the introduction of a background metric (effective quantum
field theory, string theory) gravity is simulated by gravitons
with the same space-time behaviour as photons and gluons except
for the spin and no method is know to rebuild the dynamical
chrno-geometrical structure.

On the other hand the only sufficiently developed
background-independent approach to quantum gravity, namely loop
quantum gravity, gives rise to a suggestive quantum geometry for
the 3-space but has problems with time evolution and does not
correspond to a Fock space. As a consequence, it is not known how
to introduce electro-magnetism (not to speak of the standard model
of elementary particles) so to make a comparison with ordinary
perturbative quantum electro-dynamics and understand its
modifications induced by gravity.

This state of affairs motivated an attempt \cite{2,3,4} to revisit
classic metric gravity \cite{2} and its ADM Hamiltonian
formulation to see whether it is possible to define a model of
general relativity able to incorporate fields and particles and
oriented to a background-independent quantization. First of all to
include fermions it is natural to resolve the metric tensor in
terms of cotetrad fields \cite{3,4} [$g_{\mu\nu}(x) = E^{(\alpha
)}_{\mu}(x)\, \eta_{(\alpha )(\beta )}\, E^{(\beta )}_{\nu}(x)$;
$\eta_{(\alpha )(\beta )}$ is the flat Minkowski metric in
Cartesian coordinates] and to reinterpret the gravitational field
as a {\it theory of time-like observers endowed with tetrads},
whose dynamics is controlled by the ADM action thought as a
function of the cotetrad fields. Since the standard model of
elementary particles and its extensions are a chapter of the
theory of representations of the Poincare' group on the
non-compact Minkowski space-time and we look for a Hamiltonian
decription, the mathematical pseudo-Riemannian 4-manifold $M^4$
introduced to describe space-time is assumed to be non-compact and
globally hyperbolic. This means that it admits 3+1 splittings with
foliations whose leaves are space-like Cauchy 3-surfaces assumed
diffeomorphic to $R^3$ (so that any two points on them are joined
by a unique 3-geodesic). As discussed in Ref.\cite{5} these
3-surfaces are also {\it simultaneity surfaces}, namely a
convention for the synchronization of clocks (in general different
from Einstein's one, valid in inertial systems in special
relativity). If $\tau$ is the mathematical time labeling these
3-surfaces, $\Sigma_{\tau}$, and $\vec \sigma$ are 3-coordinates
(with respect to an arbitrary observer, a centroid $x^{\mu}(\tau
)$, chosen as origin) on them, then $\sigma^A = (\tau ,\vec \sigma
)$ can be interpreted as Lorentz-scalar radar 4-coordinates and
the surfaces $\Sigma_{\tau}$ are described by embedding functions
$x^{\mu} = z^{\mu}(\tau ,\vec \sigma )$. In these coordinates the
metric is $g_{AB}(\tau ,\vec \sigma ) = z^{\mu}_A(\tau ,\vec
\sigma )\, g_{\mu\nu}(z(\tau ,\vec \sigma ))\, z^{\nu}_B(\tau
,\vec \sigma )$ [$z^{\mu}_A = \partial z^{\mu} / \partial
\sigma^A$]. Since the 3-surfaces $\Sigma_{\tau}$ are {\it equal
time} 3-spaces with all the clocks synchronized, the spatial
distance between two equal-time events will be $dl_{12} = \int^2_1
dl\, \sqrt{{}^3g_{rs}(\tau ,\vec \sigma (l))\,
{{d\sigma^r(l)}\over {dl}}\, {{d\sigma^s(l)}\over {dl}}}\,\,$
[$\vec \sigma (l)$ is a parametrization of the 3-geodesic
$\gamma_{12}$ joining the two events on $\Sigma_{\tau}$].
Moreover, by using test rays of light we can define the {\it
one-way} velocity of light between events on $\Sigma_{\tau}$.
Therefore, the Hamiltonian description has naturally built in the
tools (essentially the 3+1 splitting) to make contact with
experiments in a relativistic framework, where simultaneity is a
frame-dependent property. The manifestly covariant description
using Einstein's equations is the natural one for the search of
exact solutions, but is inadequate to describe experiments.

Other requirements \cite{3,4} on the Cauchy and simultaneity
3-surfaces $\Sigma_{\tau}$ induced by particle physics are:

i) Each $\Sigma_{\tau}$ must be a Lichnerowitz 3-manifold
\cite{6}, namely it must admit an involution so that a generalized
Fourier transform can be defined and the notion of positive and
negative frequencies can be introduced (otherwise the notion of
particle is missing like it happens in quantum field theory in
arbitrary curved space-times).

ii) Both the cotetrad fields (and the metric tensor) and the
fields of the standard model of elementary particles must belong
to the same family of suitable weighted Sobolev spaces so that
simultaneously there are no Killing vector fields on the
space-time (this avoids the cone-over-cone structure of
singularities in the space of metrics) and no Gribov ambiguity
(either gauge symmetries or gauge copies \cite{7}) in the particle
sectors; in both cases no well defined Hamiltonian description is
available.

iii) The space-time must be {\it asymptotically flat at spatial
infinity} and with boundary conditions there attained in a way
independent from the direction (like it is needed to define the
non-Abelian charges in Yang-Mills theory \cite{7}). This
eliminates the supertranslations (the obstruction to define
angular momentum in general relativity) and reduces the {\it spi
group} of asymptotic symmetries to the ADM Poincare' group. The
constant ADM Poincare' generators should become the standard
conserved Poincare' generators of the standard model of elementary
particles when gravity is turned off and the space-time becomes
the Minkowski one. As a consequence, the {\it admissible
foliations} of the space-time must have the simultaneity surfaces
$\Sigma_{\tau}$ tending in a direction-independent way to
Minkowski space-like hyper-planes at spatial infinity, where they
must be orthogonal to the ADM 4-momentum. But these are the
conditions satisfied by the Christodoulou-Klainermann space-times
\cite{8} , which are near Minkowski space-time in a norm sense.
Therefore the surfaces $\Sigma_{\tau}$ define the {\it rest frame}
of the $\tau$-slice of the universe and there are asymptotic
inertial observers to be identified with the {\it fixed stars}
(the standard origin of rotations to study the precession of
gyroscopes in space). In this class of space-times there is an
{\it asymptotic Minkowski metric} (asymptotic background), which
allows to define weak gravitational field configurations {\it
without splitting the metric} in a background one plus a
perturbation and without being a bimetric theory of gravity.

The absence of supertranslations also implies that the lapse and
shift functions have the form $N(\tau ,\vec \sigma ) = \epsilon +
n(\tau, \vec \sigma )$, $N_r(\tau ,\vec \sigma ) = n_r(\tau ,\vec
\sigma )$ [$\epsilon = +1$ with signature (+---) and $\epsilon = -
1$ with signature (-+++)], with $n(\tau ,\vec \sigma )$, $n_r(\tau
,\vec \sigma )$ tending to zero at spatial infinity.

In ADM tetrad gravity there are 16 configuration variables: the
cotetrad fields can be parametrized in terms of $n(\tau ,\vec
\sigma )$, $n_r(\tau ,\vec \sigma )$, 3 boost parameters
$\varphi_{(a)}(\tau ,\vec \sigma )$, 3 angles $\alpha_{(a)}(\tau
,\vec \sigma )$ and cotriads $e_{(a)r}(\tau ,\vec \sigma )$ on
$\Sigma_{\tau}$ [$a = 1,2,3$]. There are 14 first class
constraints $\pi_n(\tau ,\vec \sigma ) \approx 0$, $\pi_{n_r}(\tau
,\vec \sigma ) \approx 0$, $\pi_{\varphi_{(a)}}(\tau ,\vec \sigma
) \approx 0$ (the generators of local Lorentz boosts),
$M_{(a)}(\tau ,\vec \sigma ) \approx 0$ (the generators of local
rotations), $\Theta^r(\tau ,\vec \sigma ) \approx 0$ (the
generators of the changes of 3-coordinates on $\Sigma_{\tau}$) and
${\cal H}(\tau ,\vec \sigma ) \approx 0$ (the superhamiltonian
constraint).

It can be shown \cite{2,4} that the addition of the DeWitt surface
term to the Dirac Hamiltonian (needed to make the Hamiltonian
theory well defined in the non-compact case) implies that the
Hamiltonian does not vanish on the constraint surface (no frozen
reduced phase space picture in this model of general relativity)
but is proportional to the {\it weak ADM energy} (i.e. its form as
a volume integral), which governs the $\tau$-evolution \cite{9}.
Moreover the Hamiltonian gauge transformations generated by the
superhamiltonian constraint do not have the Wheeler-DeWitt
interpretation (evolution in local time), but transform an
admissible 3+1 splitting into another admissible one (so that all
the admissible notions of simultaneity are gauge equivalent). See
Refs.\cite{2,4} for the scheme of addition of gauge fixings and,
in particular, for the bibliography showing that a completely
fixed Hamiltonian gauge becomes a unique 4-coordinate system on
space-time only on the solutions of Einstein's equations
(on-shell), so that it describes an extended physical laboratory.
Finally it can be shown that Lichnerowicz's identification of the
conformal factor of the 3-metric on $\Sigma_{\tau}$ ($\phi =
[det\, {}^3g]^{1/12}$) as the unknown in the superhamiltonian
constraint is the correct one: as a consequence the gauge variable
describing the normal deformations of the simultaneity surfaces
$\Sigma_{\tau}$ is the momentum $\pi_{\phi}(\tau ,\vec \sigma )$
canonically conjugate to $\phi (\tau ,\vec \sigma )$ (and not the
trace of the extrinsic curvature of $\Sigma_{\tau}$, the so called
intrinsic York time).

In Ref.\cite{10} there is a review of the various implications of
Einstein's Hole Argument,namely of the property of general
covariance. The invariance of Einstein's equations under active
diffeomorphisms (the widest local symmetry group of general
relativity, whose passive reinterpretation by Bergmann and Komar
contains the ordinary coordinate transformations and the Legendre
pull-back of the Hamiltonian gauge transformations) imply: i) the
absence of determinism (only two of Einstein's equations contain
dynamical information: four are restrictions on initial data and
four are void due to Bianchi identities), i.e. the presence of
arbitrary gauge variables; ii) absence of a physical individuation
of the mathematical points as physical point-events of space-time.
On one side such an individuation can be achieved by formulating
four of the gauge fixing constraints as the requirement that a set
of ordinary 4-coordinates coincides with four suitable scalar
functions of the four eigenvalues of the Weyl tensor
(Bergmann-Komar intrinsic pseudo-coordinates). This implies that
in a sense  the {\it space-time is the gravitational field itself}
and that each 4-coordinate system has a {\it noncommutative
structure} associated to it already at the classical level.

On the other side, following Dirac, we have to identify a
canonical basis $r_{\bar a}(\tau ,\vec \sigma )$, $\pi_{\bar
a}(\tau ,\vec \sigma )$, $\bar a = 1,2$, of {\it Dirac
observables} (DO) as the physical degrees of freedom of the
gravitational field, i.e. a canonical basis of predictable
gauge-invariant quantities satisfying deterministic Hamilton
equations governed by the weak ADM energy. This can be achieved by
means of a Shanmugadhasan canonical transformation adapted to 13
of the 14 first class constraints (not to the superhamiltonian
one), which turns out to be a {\it point} canonical transformation
as a consequence of the form of the finite gauge transformations.
In the new canonical basis 13 new momenta vanish due to the 13
constraints and their 13 conjugate configuration variables are
{\it abelianized gauge variables} (as already said the 14th one is
$\pi_{\phi}$). While the 14 gauge variables describe {\it
generalized inertial effects}, the DO describe {\it generalized
tidal effects}. Since the transformation is a point one, the old
momenta are linear functionals of the new ones, with the kernels
determined by a set of elliptic partial differential equations. In
this way for the first time we can identify a canonical basis of
non-local and non-tensorial DO, which remains canonical in the
class of gauges where $\pi_{\phi}(\tau ,\vec \sigma ) \approx 0$,
even if no one knows how to solve the superhamiltonian constraint,
i.e. the Lichnerowicz equation for the conformal factor.

A special completely fixed Hamiltonian gauge in this class can be
defined by adding other 13 suitable gauge fixing constraints. In
this gauge, which turns out to be non-harmonic in the weak field
regime, the 3-metric on $\Sigma_{\tau}$ is {\it diagonal} and it
corresponds to a unique 3-orthogonal 4-coordinate system on
space-time (with an associated admissible 3+1 splitting with well
defined simultaneity leaves) on the solutions of Hamilton
equations. In this gauge it is possible to give a {\it
background-independent} definition of a weak gravitational field:
the DO $r_{\bar a}(\tau ,\vec \sigma )$, $\pi_{\bar a}(\tau ,\vec
\sigma )$ should be slowly varying on a wavelength of the
resulting post-Minkowskian gravitational wave, with the
configurational DO $r_{\bar a}$ replacing the two polarizations of
the harmonic gauges. A {\it Hamiltonian linearization} is defined
in the following way \cite{11}:

i) Assuming $ln\, \phi (\tau ,\vec \sigma ) = O(r_{\bar a})$, the
Lichnerowitz equation can be linearized and for the first time a
non-trivial solution for $\phi$ can be found. Using this solution
all the other constraints and the elliptic canonicity conditions
can be linearized and solved. By putting these solutions in the
integrand of the weak ADM energy, we get a well defined form for
the energy density in this gauge in terms of the DO, i.e. the
physical degrees of freedom of the gravitational field.

ii) The resulting ADM energy is approximated with the terms {\it
quadratic} in the DO and the resulting linearized Hamilton
equations are studied and solved. It is explicitly checked that
the linerized Einstein's equations are satisfied by this solution.
Even if the gauge is not harmonic, the wave equation $\Box r_{\bar
a}(\tau ,\vec \sigma) = 0$ is implied by the Hamilton equations
and solutions satisfying the universe rest-frame condition are
found (they cannot be transverse waves in the rest frame). These
are the {\it post-Minkowskian background-independent gravitational
waves}. The deformation patterns of a sphere of test particles
induced by $r_{\bar 1}$ and $r_{\bar 2}$ are determined by
studying the geodesic deviation equation.

We are now studying tetrad gravity coupled to a perfect fluid
described by a suitable singular Lagrangian \cite{12}. The
Hamiltonian linearization in the special 3-orthogonal gauge,
together with an adaptation to our formalism of the theory of
Dixon's multipoles \cite{13}, will allow to find the
post-Minkowskian (without any post-Newtonian approximation!)
generalization of the quadrupole emission formula and the explicit
form in this gauge of the action-at-a-distance Newton and
gravito-magnetic potentials inside the fluid together with its
tidal interactions. The resulting formalism should help to find a
description of binary systems in a post-Minkowskian regime, where
the post-Newtonian approximations fail. Moreover, the two-body
problem of general relativity in the post-Minkowskian weak field
regime will be studied by using a new semi-classical
regularization of the self-energies, implying the $i \not= j$ rule
like it appens in the electro-magnetic case \cite{14}. Also tetrad
gravity coupled to Klein-Gordon, electro-magnetic and Dirac fields
is under investigation.

It will be explored the possibility of defining a scheme of
Hamiltonian numerical relativity, based on expansions in the
Newton constant G (the so called post-Minkowskian approximations),
to study the strong field regime of tetrad gravity.

Finally we will try to find the Hamiltonian re-formulation of the
Newman-Penrose formalism. This would allow to look for a
Shanmugadhasan canonical basis in which the physical degrees of
freedom of the gravitational field are described by {\it Bergmann
observables} (BO), namely a set of coordinate-independent scalar
DO. Such a basis would allow to start a new program of
quantization of gravity, based on the idea of quantizing only the
tidal effects (the BO) and not the inertial ones (the gauge
variables), since the latter describe only the {\it appearances}
of the phenomena. A prototype of this quantization is under study
in special relativity to arrive at a formulation of atomic physics
in non-inertial systems: while for relativistic particles (and
their non-relativistic limit) there are already preliminary
results \cite{15}, for the inclusion of the electro-magnetic field
we have to find a way out from the Torre-Varadarajan no-go
theorem, the obstruction to the Tomonaga-Schwinger formalism.
Moreover, if the weak ADM energy in a completely fixed Hamiltonian
gauge can be expressed in terms of BO, this would help to clarify
the problem of the coordinate-dependence of the energy density in
general relativity, which we think is a preliminary step for a
correct understanding of the cosmological constant and dark energy
problems.

\end{document}